\documentclass[aps, prb, preprint, amsmath,amssymb,showpacs,superscriptaddress, longbibliography]{revtex4-1}
\usepackage{graphicx}
\usepackage{color}
\definecolor{green}{RGB}{34,139,34}
\makeatletter

\begin{document}%

\title{Emergence of  Fermi arcs and novel magnetic splitting in an antiferromagnet}

\author{Benjamin Schrunk*}
\affiliation{Ames Laboratory, Ames, Iowa 50011, USA}

\author{Yevhen Kushnirenko*}
\affiliation{Ames Laboratory, Ames, Iowa 50011, USA}
\affiliation{Department of Physics and Astronomy, Iowa State University, Ames, Iowa 50011, USA}

\author{Brinda Kuthanazhi}
\affiliation{Ames Laboratory, Ames, Iowa 50011, USA}
\affiliation{Department of Physics and Astronomy, Iowa State University, Ames, Iowa 50011, USA}

\author{Junyeong Ahn}
\affiliation{Department  of  Physics,  Harvard  University,  Cambridge  MA  02138,  USA}

\author{Lin-Lin Wang}
\affiliation{Ames Laboratory, Ames, Iowa 50011, USA}

\author{Evan O'Leary}
\affiliation{Ames Laboratory, Ames, Iowa 50011, USA}
\affiliation{Department of Physics and Astronomy, Iowa State University, Ames, Iowa 50011, USA}

\author{Kyungchan Lee}
\affiliation{Ames Laboratory, Ames, Iowa 50011, USA}
\affiliation{Department of Physics and Astronomy, Iowa State University, Ames, Iowa 50011, USA}

\author{Andrew Eaton}
\affiliation{Ames Laboratory, Ames, Iowa 50011, USA}
\affiliation{Department of Physics and Astronomy, Iowa State University, Ames, Iowa 50011, USA}

\author{Alexander Fedorov}
\affiliation{Leibniz Institute for Solid State and Materials  Research, 01069 Dresden, Germany}
\affiliation{Helmholtz-Zentrum Berlin f\"ur Materialien und Energie, Berlin, Germany}

\author{Rui Lou}

\affiliation{School of Physical Science and Technology, Lanzhou University, Lanzhou 730000, China}

\affiliation{Leibniz Institute for Solid State and Materials  Research, 01069 Dresden, Germany}

\author{Vladimir Voroshnin}
\affiliation{Helmholtz-Zentrum Berlin f\"ur Materialien und Energie, Berlin, Germany}

\author{Oliver J. Clark}
\affiliation{Helmholtz-Zentrum Berlin f\"ur Materialien und Energie, Berlin, Germany}

\author{Jaime S\'anchez-Barriga}
\affiliation{Helmholtz-Zentrum Berlin f\"ur Materialien und Energie, Berlin, Germany}

\author{Sergey L. Bud'ko}
\affiliation{Ames Laboratory, Ames, Iowa 50011, USA}
\affiliation{Department of Physics and Astronomy, Iowa State University, Ames, Iowa 50011, USA}

\author{Robert-Jan Slager}
\email[]{rjs269@cam.ac.uk}
\affiliation{Department  of  Physics,  Harvard  University,  Cambridge  MA  02138,  USA}
\affiliation{TCM Group, Cavendish Laboratory, University of Cambridge, Cambridge CB3 0HE, United Kingdom}

\author{Paul C. Canfield}
\email[]{canfield@ ameslab.gov}
\affiliation{Ames Laboratory, Ames, Iowa 50011, USA}
\affiliation{Department of Physics and Astronomy, Iowa State University, Ames, Iowa 50011, USA}

\author{Adam Kaminski}
\email[]{adamkam@ameslab.gov; * - these authors contributed equally}
\affiliation{Ames Laboratory, Ames, Iowa 50011, USA}
\affiliation{Department of Physics and Astronomy, Iowa State University, Ames, Iowa 50011, USA}

\date{\today}
\maketitle 
{\bf  }

{\bf The Fermi arcs are signatures of exotic states in solids because they defy conventional concept of Fermi surfaces as closed contours in momentum space. Fermi arcs were first discovered in cuprates, and caused by the pseudogap\cite{Mike_Nature1998}. Weyl semimetals provided another way to generate Fermi arcs by breaking either the time reversal symmetry (TRS)~\cite{Wang2016Time} or inversion symmetry~\cite{Xu15SciDis} of a 3D Dirac semimetal, which can result in a Weyl semimetal with pairs of Weyl nodes that have opposite chirality\cite{Armitage2018Weyl}. The bulk-boundary correspondence associated with the Chern number leads to the emergence of Fermi arcs on the boundary \cite{Xu15SciDis,Huang2016Spectroscopic,Tamai2016Fermi,Deng2016Experimental,Ashvin_PRB2011,Hirschberger_NMAT2016,Felser_Science2019, Felser_PRB2018}. Here, we present experimental evidence that pairs of magnetically split hole- and electron-like Fermi arcs emerge below the Neel temperature (T$_N$), in the antiferromagnetic (AFM) state of cubic NdBi  due to a novel band splitting effect. Whereas TRS is broken by the AFM order, both inversion and nonsymmorphic TRS are preserved in the bulk, precluding the possibility of a Weyl semimetal. The observed magnetic splitting is highly unusual, as it creates bands of opposing curvature, that changes with temperature and follows the antiferromagnetic order parameter. This is completely different from previously reported cases of magnetic splittings such as traditional Zeeman and Rashba, where the curvature of the bands is preserved.
Therefore, our finding represents a new Fermionic state created by new type of magnetic band splitting in the presence of a long-range AFM order that are not readily explained by existing theoretical ideas. }

Rare-earth monopnictides \cite{KumigashiraPRB1996,Paul_Phil1992, Paul_Jcryst2001, Yun_PRB2017,Kuroda_2018} were recently proposed to host novel topological states \cite{DuanCommPhys2018,GuoNPJ2017,ZhuPRB2020}, and studied experimentally \cite{Kuroda_2018,Kuroda_2020,matt2020}. The bulk Fermi surface of cubic NdBi (Fig. 1a) is similar to other rare earth monopnictides \cite{Yun_PRB2017,Kuroda_2018,DuanCommPhys2018} and consists of several hole pockets at the center of the Brillouin zone and ellipsoidal electron pockets in each corner of the zone as shown in Fig. 1b, c. Magnetic susceptibility M/H as a function of temperature is shown in Fig. 1d. A clear maximum is seen around 24 K indicating the onset of magnetic ordering below that temperature, which leads to the clear loss of spin disorder scattering below this temperature in the electrical resistivity shown in Fig. 1e. The high temperature M(T)/H(T) data follow a Curie-Weiss behavior with Weiss temperature $\theta$ = 17.2 $\pm$ 0.1~K and an effective moment $\mu$ = (3.2 $\pm$ 0.1) $\mu_B$, which is close to the expected value of 3.6 $\mu_B$ for a Hund's rule ground state multiplet, J=9/2. The inset of Fig. 1d shows the magnetization as a function of applied fields measured at 2 K. A linear behavior is observed up to applied fields of 50 kOe. Both M/H(T) and M(H) behavior indicate an AFM ordering below T$_N$ = 24 K.  

The measured Fermi surface in the paramagnetic (PM) state is shown in Fig. 2a, and the dispersion along selected cuts is shown in panels Fig. 2b, c. This is in good agreement with bulk band structure calculations (Fig. 1b, c). There are very broad areas of intensity in the proximity of E$_F$, consistent with the mainly 3D character of the bulk bands. The lack of sharp features indicates that there are no significant presence of surface states near E$_F$ in the paramagnetic state. 

Upon cooling below T$_N$, a long-range AFM order develops\cite{Tsuchida_1965, nereson_1971,Schobinger_1973} and a new, very sharp features appear near E$_F$ that are consistent with the emergence of surface states as seen in Fig. 2d-f. A new set of electron pockets appear, one residing within the bulk electron pocket and the other near the corners of the bulk hole pockets. Even more interestingly, four new, very sharp, disconnected contours appear around the center of the zone. It should be noted that these features appear only in the AFM state, and there is not even a hint of their intensity in these parts of the momentum space above T$_N$. These features are induced by the presence of the long-range AFM order and are not artifacts of the sharpening due to the enhancement of quasiparticle lifetime. The spectral intensity of these features appears only below T$_N$ and is increasing upon cooling. In fact, as discussed below, the band splitting and weight of these features follow order-parameter-like temperature dependencies below T$_N$. The Fermi surface and band structure are fully 4-fold rotationally symmetric, and the apparent broadening of the arcs located along horizontal direction is due to larger data bins in this direction. 

We examine the structure of the Fermi arcs by plotting the detailed dispersion along several cuts in the momentum space in Fig. 3. The band that gives rise to the Fermi arc displays a sharp dispersion over a relatively wide energy range close to the symmetry direction crossing the center of the zone shown in Fig. 3b. As we move away from symmetry direction, the energy range over which this band has sharp intensity decreases, and becomes limited to few tens of meV near the tips of the arcs. Beyond the tips of the arc (Fig. 3b \#4) the sharp surface state intensity vanishes and only very broad 3D bulk band intensity is present (Fig. 3b \#5 and \#6). Although graphically the arcs may seem reminiscent of cuprates, the situation is completely different. In cuprates the band giving rise to Fermi surface is present at all temperatures and the pseudogap suppresses the spectral weight of the band only near E$_F$, while leaving the higher binding energy portion unaffected\cite{Mike_Nature1998}, which is clearly not the case in NdBi. 

In order to verify the intrinsic nature of Fermi arcs discussed above we need to exclude a possibility that portion of the Fermi surface is suppressed by polarization selection rules and matrix elements. We do this by performing measurements on the same cleaved surface after rotating the sample by 45 degrees. The Fermi arcs are still present in such obtained data at the same locations as before as shown in Fig. 3c, d. Detailed cuts provide further evidence that the band that gives rise to Fermi arcs merges with bulk bands, as shown in Fig. 3d. These data also reveal that there is a relation between the Fermi arc band and the surface state electron pockets that exist below T$_N$ as they appear to disperse together.

Perhaps the most remarkable aspect of surface states described above is their formation below T$_N$ and temperature evolution. To illustrate this we plot in Fig. 4a, the Fermi surface and in Fig. 4b band dispersion along selected cut for several temperatures below and above T$_N$. The data at 30~K and 25~K reveal only broad 3D bands and the surface states are completely absent. At 24~K, a hint of intensity appears in the band dispersion (indicated by red arrow). At 23K, there is clearly visible Fermi arc in the FS plot and slightly broaden linear dispersion. At 20~K, this Fermi arc splits into two, each having opposing curvature. The dispersion data reveals splitting of initial surface state band into hole-like and electron like bands. Their intensity vanishes away from $\Gamma$, which forms the Fermi arcs. At 14K the separation between the two bands increases, mostly due to the upward movement of the electron-like band and rapid change of its curvature. At the same time its intensity near E$_F$ increases and the electron-like Fermi arc transforms into an electron pocket, area of which decreases with temperature (see two bottom panels of Fig. 4a. 

We quantify the temperature evolution of the surface states by plotting the EDC's at a momentum near the bottom of the electron band as indicated by a dashed line in the bottom panel of Fig. 4b. Above T$_N$, the EDC is a linear intensity cut-off by the Fermi function due to the projection of bulk bands. Just below T$_N$, a small peak appears at $\sim$70 meV. Its intensity is increasing upon cooling and a few degrees below T$_N$, its splits into two peaks. The peak closer to E$_F$ is due to an electron-like band and the other one is due to a hole-like band. Upon further cooling, the ``hole" peak remains mostly at the same energy of $\sim$85~meV (marked by dotted line in Fig. 4c), while increasing in intensity. The electron peak moves closer to E$_F$ and resides at $\sim$38 meV at 5.5~K.  In Fig. 4d we plot the area of both peaks after subtracting and normalizing by the EDC above T$_N$. In Fig. 4e we plot the energy separation between the EDC peaks as a function of temperature. In both cases those quantities behave in a manner similar to AFM order parameter.

We would like to emphasize the unusual nature of magnetic splitting we report here. In the regular Zeeman interaction, such as in an itinerant ferromagnet, presence of an internal magnetic field splits degenerate bands into minority and majority bands and rigidly shifts them up and down in energy as illustrated on the left side of Fig. 4f \cite{Himpsel}. In the Rashba interaction, the spin momentum locking causes the degenerate band to split in a way that makes the products appear rigidly shifted along the momentum axis \cite{Rashba}, as illustrated on the right side of in Fig. 4f. In the case reported here, a linearly dispersing surface state that appears just below T$_N$ is split into two bands with opposed curvature: electron-like  and hole-like as illustrated in Fig. 4g. Upon cooling and thus increase of internal magnetic field, the hole band remains mostly unchanged, while the curvature of the electron band increases as it moves to lower binding energies (Fig. 4g). This behavior is incompatible with known cases of magnetic splitting or presence of energy gap, as the latter would mostly affect adjacent portions of the bands.

The opposite band curvature of two bands and the sign-changing pattern of the dichroic response indicate that the surface states are massive Dirac fermions.
Therefore, a possible scenario is that the surface spectrum features topological surface states. 
In this scenario, there are massless surface Dirac points protected by in particular $C_{2z}T$ symmetry in the paramagnetic state, where $C_{2z}$ is the 180$^{\circ}$ rotation around the surface normal direction, but they are masked because their spectrum overlap with the bulk spectrum.
A pair of split electron-like and hole-like bands is then generated by the gap opening of each Dirac point due to $C_{2z}T$ breaking below T$_N$.
However, this does not explain how the gap opening enhances the spectral weight on the surface.
More importantly, DFT calculations do not corroborate this scenario.
Other ideas like the floating bands~\cite{topp2017surface} due to broken non-symmorphic time reversal symmetry on the surface are not supported by DFT calculations either.

In fact, our DFT calculations do not reproduce the experimentally observed band structure with the reported A-type AFM (AFMA)~\cite{Tsuchida_1965,nereson_1971}, implying that any simple attempts based on band theory will fail.
 In Supplementary Information, we present the band structure and Fermi surfaces for AFMA (shown in Fig.1a) and also AFMC (intralayer checkerboard with out-of-plane moment) orders. No sign of surface states is found at the Fermi level  for AFMA around the zone center. While surface states that are absent in the non-magnetic state emerges for AFMC at 150 meV below $E_F$, the overall band structure looks very different from the experimental data because AFMC clearly breaks $C_4$ symmetry.
While the mismatch between our experiment and DFT calculations implies that the correlation effect can play an important role, we do not observe a sign of the Kondo effect, which further underpins the puzzling nature of the observed surface states.
Our observation suggests the existence of surface-resonant states in the continuum of bulk states, forming disconnected arcs at the Fermi level. However, the robustness of the surface resonance against the hybridization with the bulk bands is yet to be explained.

In any case, it is evident that magnetism and spin-orbit coupling are very important. As such, the delicate interplay with magnetism, spin-orbit coupling, and potential topology underlying the emergence of the Fermi arcs should provide a fascinating theory impetus, where all symmetries and specific processes need to be taken into account. Because the surface states we observe are intimately linked to the AFM order, they can be likely controlled by the application of magnetic fields as well as temperature pulses such as those present in ultrafast laser pulses, thereby opening new avenues for terahertz spintronics.

\section*{References}

\bibliography{NdBiarcs}

\clearpage
\section*{methods}

\noindent{{\bf Single crystal growth of NdBi.} Single crystals of NdBi were grown out of a Bi rich, binary melt. The elements, with an initial stoichiometry of Nd14Bi86, were put into a fritted alumina crucible \cite{Canfield2016Use} (sold by LSP Ceramics as Canfield Crucible Sets \cite{FrittURL}) and sealed in a fused silica tube under a partial pressure of argon. The thus prepared ampoule was heated up to 1100$^\circ$~C over 5 hours and held there for 5 hours. This was followed by a slow cooling to 680$^\circ$~C over 80 hours and decanting of the excess liquid using a centrifuge \cite{Canfield2020RPP}. The cubic crystals obtained were stored and handled in a glovebox under a Nitrogen atmosphere. The magnetic measurements were carried out in a Quantum Design Magnetic Property Measurement System with applied magnetic fields up to 5 T. 
The ac resistivity measurements were done using the ACT option in a Quantum Design Physical Property Measurement System. The measurement was done in the standard four point configuration and contacts were made using silver paint Dupont 4929, with contact resistance values of about 4-6 ohms. " }

\noindent{{\bf ARPES measurements.} ARPES data was collected using vacuum ultravilet (VUV) laser ARPES spectrometer that consists of a Scienta DA30 electron analyzer, picosecond Ti:Sapphire oscillator and fourth-harmonic generator \cite{jiang2014tunable}. Data from the laser based ARPES were collected with 6.7~eV photon energy. Angular resolution was set at $\sim$ 0.1$^{\circ}$ and 1$^{\circ}$, along and perpendicular to the direction of the analyser slit respectively, and the energy resolution was set at 2 meV. The VUV laser beam was set to vertical polarization, i. e. along k'$_y$ direction. The diameter of the photon beam on the sample was $\sim 20\,\mu$m. Samples were cleaved \textit{in-situ} at a base pressure lower than 2$\times$ 10$^{-11}$ Torr. 
Results were reproduced using several different single crystals and temperature cycling. }


\noindent{{\bf DFT calculations.} For the bulk band structure of non-magnetic NdBi calculated in Fig.1, we used PBE exchange-correlation functional in DFT including the spin-orbit coupling (SOC) with the 4\textit{f} orbitals treated as core electrons as implemented in VASP. The experimental lattice constant of 6.41 \AA, a \textit{$\Gamma$}-centered Monkhorst-Pack (13$\times$13$\times$13) \textit{k}-point mesh with a Gaussian smearing of 0.05 eV and a kinetic energy of 183 eV have been used. The DFT calculation for non-magnetic NdBi gives a band inversion around three \textit{X} points and the associated topological surface states for a strong topological insulator, which are similar to those features in LaBi. However, ARPES data shows the absence of such surface states in the paramagnetic NdBi above T$_N$.}

\section*{Data availability}
Relevant data for the work are available at the Materials Data Facility Ref. XX

\section*{Acknowledgements}
We would like to thank Ashvin Vishwanath and J\"org Schmalian for useful discussions and comments. ARPES measurements were supported by the U.S. Department of Energy, Office of Basic Energy Sciences, Division of Materials Science and Engineering. Ames Laboratory is operated for the U.S. Department of Energy by Iowa State University under Contract No. DE-AC02-07CH11358. Crystal growth and characterization were supported by Center for the Advancement of Topological Semimetals (CATS), an Energy Frontier Research Center funded by the U.S. DOE, Office of Basic Energy Sciences.  R.-J.~S.~acknowledges funding from the Marie Sk{\l}odowska-Curie programme under EC Grant No. 842901 and the Winton programme as well as Trinity College at the University of Cambridge. J.A. was supported by the Basic Science Research Program through the National Research Foundation of Korea (NRF) funded by the Ministry of Education (Grant No. 2020R1A6A3A03037129) and CATS. A.F. and R.L acknowledge support from SFB1143 ``Correlated Magnetism" W\"urzburg-Dresden Cluster of Excellence on Complexity and Topology in Quantum Matter-ct.qmat. R.L acknowledges support from 
National Natural Science Foundation of China (Grant No. 11904144). J. S.-B. gratefully acknowledges financial support from the Impuls- und Vernetzungsfonds der Helmholtz-Gemeinschaft under grant No. HRSF-0067 (Helmholtz-Russia Joint Research Group).

\section*{Author contributions}
B. S., P. C. C. and A. K. conceived and designed the experiment. Y. K. and A. K. performed analysis of ARPES data. J. A. and R.-J. S provided theoretical analysis and interpretation. B.K., S. L. B. and P.C.C. grew and characterized the samples. L.-L. W. performed DFT calculations. B. S., Y. K., E. O., K. L., A. E., A. F., R. L., V. V., O. J. C., J. S.-B and A.K. performed ARPES measurements and support. The manuscript was drafted by J. A., B. S., R.-J. S., P. C. C. and A. K. All authors discussed and commented on the manuscript.

\section*{Competing interests}
The authors declare no competing interests.

\begin{figure*}[!ht]
    \includegraphics[scale=0.5] {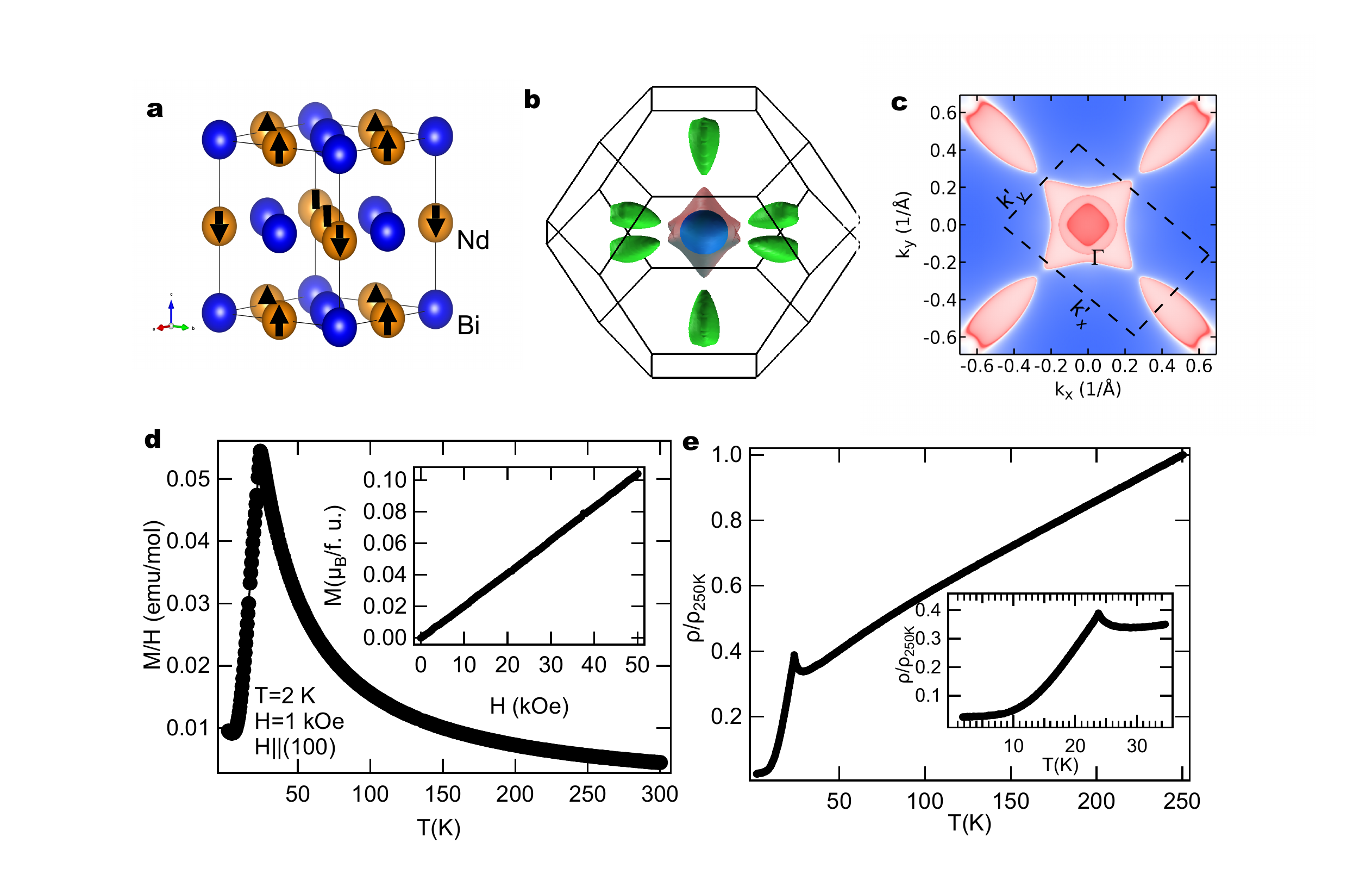}
	\caption{{\bf Structure, Fermi surface and magnetic properties of NdBi crystals.}
	{\bf a,} Crystal structure of NdBi.
	{\bf b,} DFT-calculated non-magnetic 3D Fermi surface
	{\bf c,} 2D projection of bulk Fermi surface on (001) surface showing two hole sheets and one electron sheet at the zone center, and another electron pocket towards the corners of the Brillouin zone. Dashed rectangle marks area where data for Fig. 2 and 3 were measured. 
	{\bf d,} Temperature dependence of magnetization of NdBi measured in field of 1 kOe. Inset shows the field dependence of magnetization at T=2~K.
	{\bf e,} Temperature dependence of resistivity normalized by value at 250~K. Inset shows temperature range close to T$_N$.
	\label{fig:fig1}}
\end{figure*}

\begin{figure*}[!htb]
    \includegraphics[scale=0.65] {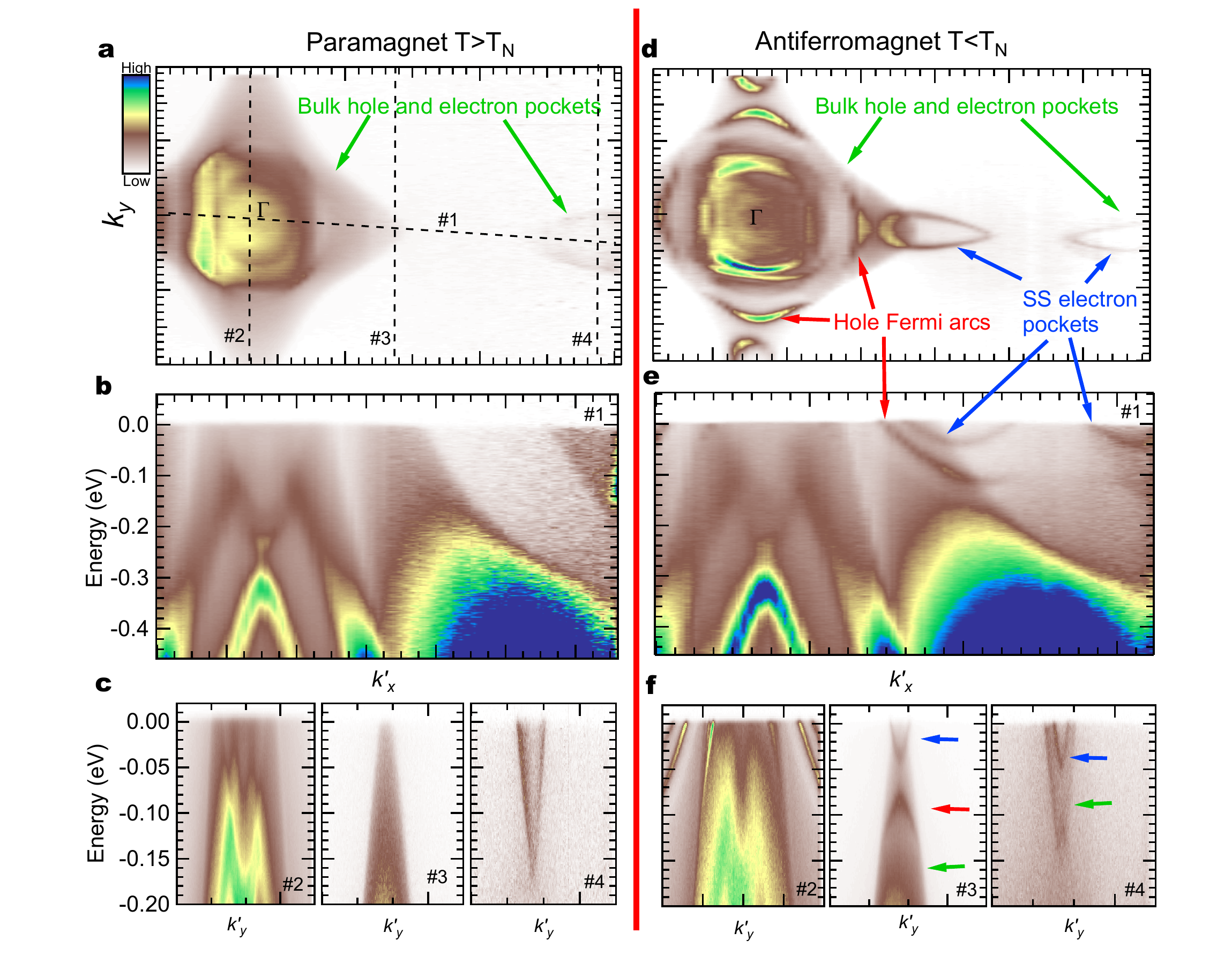}
	\caption{{\bf Fermi surface and band dispersion in paramagnetic and AFM states of NdBi measured within dashed rectangle marked in Fig. 1c.}
	{\bf a,} Plot of ARPES intensity  integrated within $\pm$5 meV about E$_F$ at T=30~K, in the paramagnetic state. High intensity regions mark locations of Fermi surface. Color scale shown is used in all other image plots.
	{\bf b,} and {\bf c,} measured band dispersion along several directions marked by dashed lines in ({\bf a}).
	{\bf d,} Plot of ARPES intensity intergrated within $\pm$5 meV about E$_F$ at T=6~K, in the AFM state. High intensity regions mark locations of Fermi surface. Location of surface state Fermi arcs and electron pockets are marked by arrows.
	{\bf e,} and {\bf f,} band dispersion measured in AFM state at T=~6K along the same directions as in ({\bf b, c}) marked by dashed lines in ({\bf a}). Blue and red arrows in cut \#3 point to surface state electron and hole bands respectively. Green arrow points to the bulk band.
	\label{fig:fig2}}
\end{figure*}

\begin{figure}[!htb]
    \includegraphics[scale=0.65]{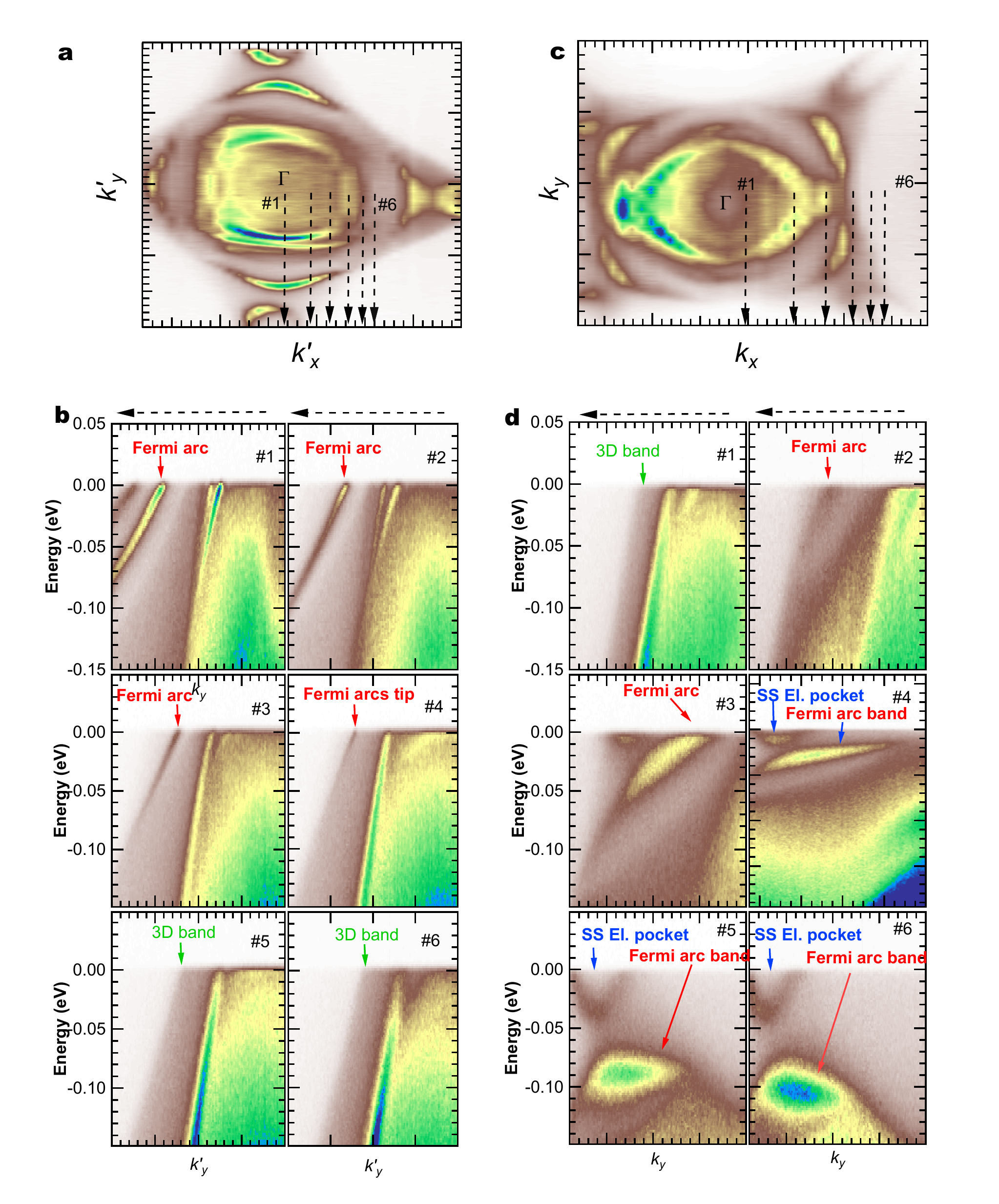}
	\caption{{\bf Dispersion of Fermi arc bands.}
	{\bf a,} Fermi surface map near $\Gamma$ in AFM state at T=6~K.
	{\bf b,} Band dispersion along cuts marked by dashed arrows in ({\bf a}).
	{\bf c,} Fermi surface map near $\Gamma$ in AFM state at T=6~K after rotating the sample by 45~deg.
	{\bf d,} Band dispersion along cuts marked by arrows in ({\bf c}).
	\label{fig:fig3}}
\end{figure}

\begin{figure}[!htb]
    \includegraphics[scale=0.4]{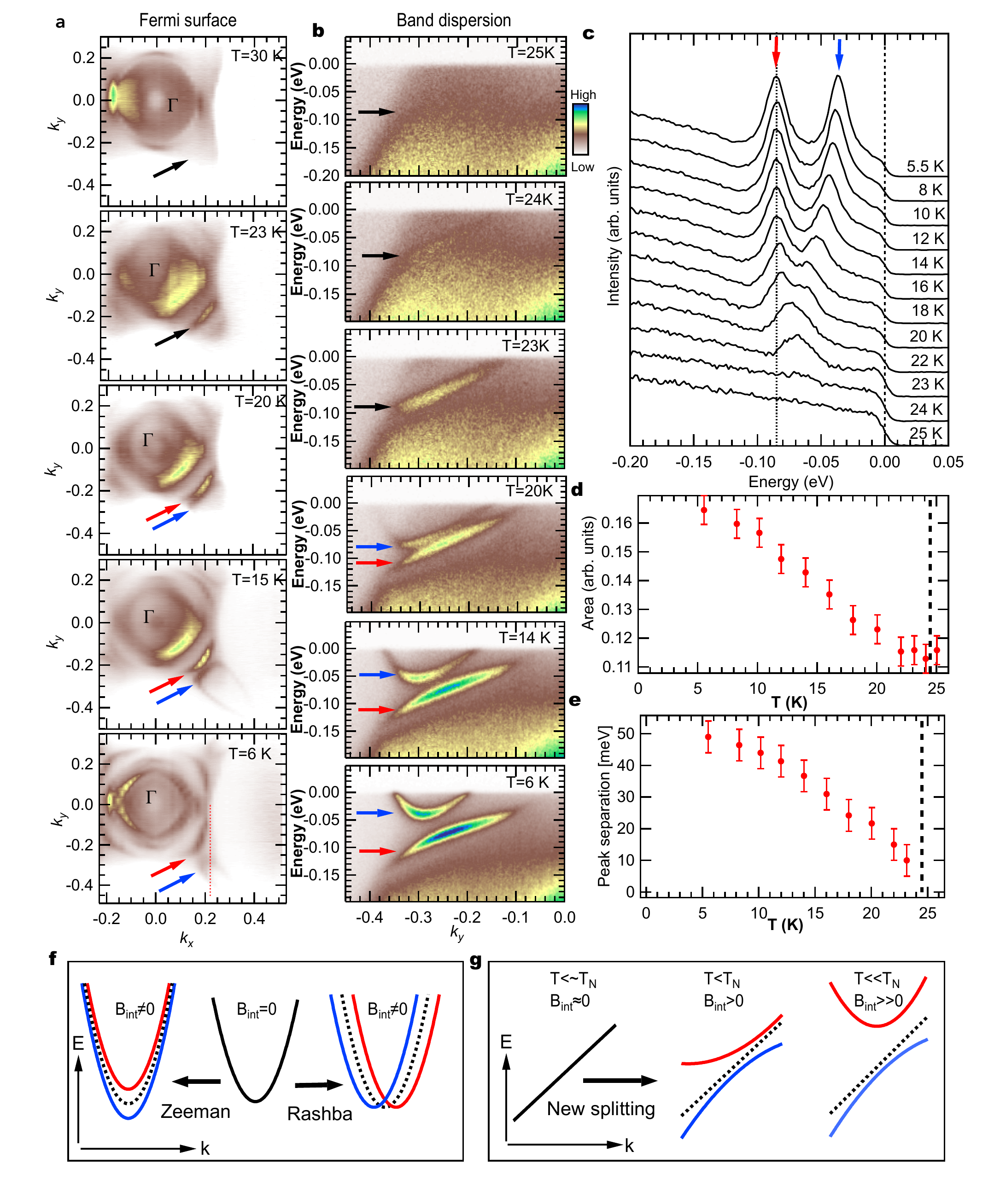}
	\caption{{\bf Temperature dependence of the Fermi arcs. }
	{\bf a,} Fermi surface map near $\Gamma$ for several temperatures below and above T$_N$. Black arrows mark area where surface state appears just below T$_N$. Red and blue arrows point to split hole- and electron-like Fermi arcs/pockets respectively.
	{\bf b,} Band dispersion along cuts marked by red dashed line in bottom panel in ({\bf a}). Upper band forms electron-like Fermi arc at temperatures slightly below T$_N$=24.5K, which expands to a closed, electron FS sheet at very low temperatures. Black arrows mark area where surface state appears just below T$_N$. Red and blue arrows point to split hole- and electron-like bands  respectively.
	{\bf c,} EDC's at momentum marked by black dashed line in bottom panel of ({\bf b}) normalized at binding energy of -0.15~eV. Red and blue arrows mark EDC peaks from hole- and electron-like bands. Dotted line marks binding energy of EDC peak for hole band at low temperature. 
	{\bf d,} Plot of the area of EDCs in panel ({\bf c}) integrated between -0.15~eV and 0.05~eV. Dashed vertical line marks value of T$_N$.
	{\bf e,} Plot of energy separation of EDC peaks from panel ({\bf c}). Dashed vertical line marks value of T$_N$.
	{\bf f,} Schematic illustration of regular Zeeman and Rashba band splittings.
	{\bf g,} Schematic illustration of novel band splitting observed in our data.
	\label{fig:fig4}}
\end{figure}

\newpage
\clearpage

\begin{center}
{\bf Supplementary Information for ``Emergence of Fermi arcs due to  magnetic band splitting in an antiferromagnet"}
\end{center}

\newpage
\clearpage
\renewcommand{\figurename}{Supplementary Fig.}
\setcounter{figure}{0}

\section{Sample characterization}

Supplementary Fig. S1 compares the derivatives of resistivity and magnetization (d$\rho$/dT and d(M(T)/H)/dT) as a function of temperature \cite{Fisher1962,Fisher1968}. These data show reasonable agreement between the two measurements with a transition temperature of T$_N$ = 23.9~K. 

NdBi samples were characterized by powder x-ray diffraction using a Rigaku Miniflex diffractometer, in a glove box under nitrogen atmosphere, with Cu K$\alpha$ radiation. The diffraction pattern confirmed the phase purity, and the NaCl-type structure in the cubic space group Fm-3m, agreeing with the reported structure as shown in Supplementary Fig. S2. 

\begin{figure*}[tb]
	\includegraphics[width=6 in]{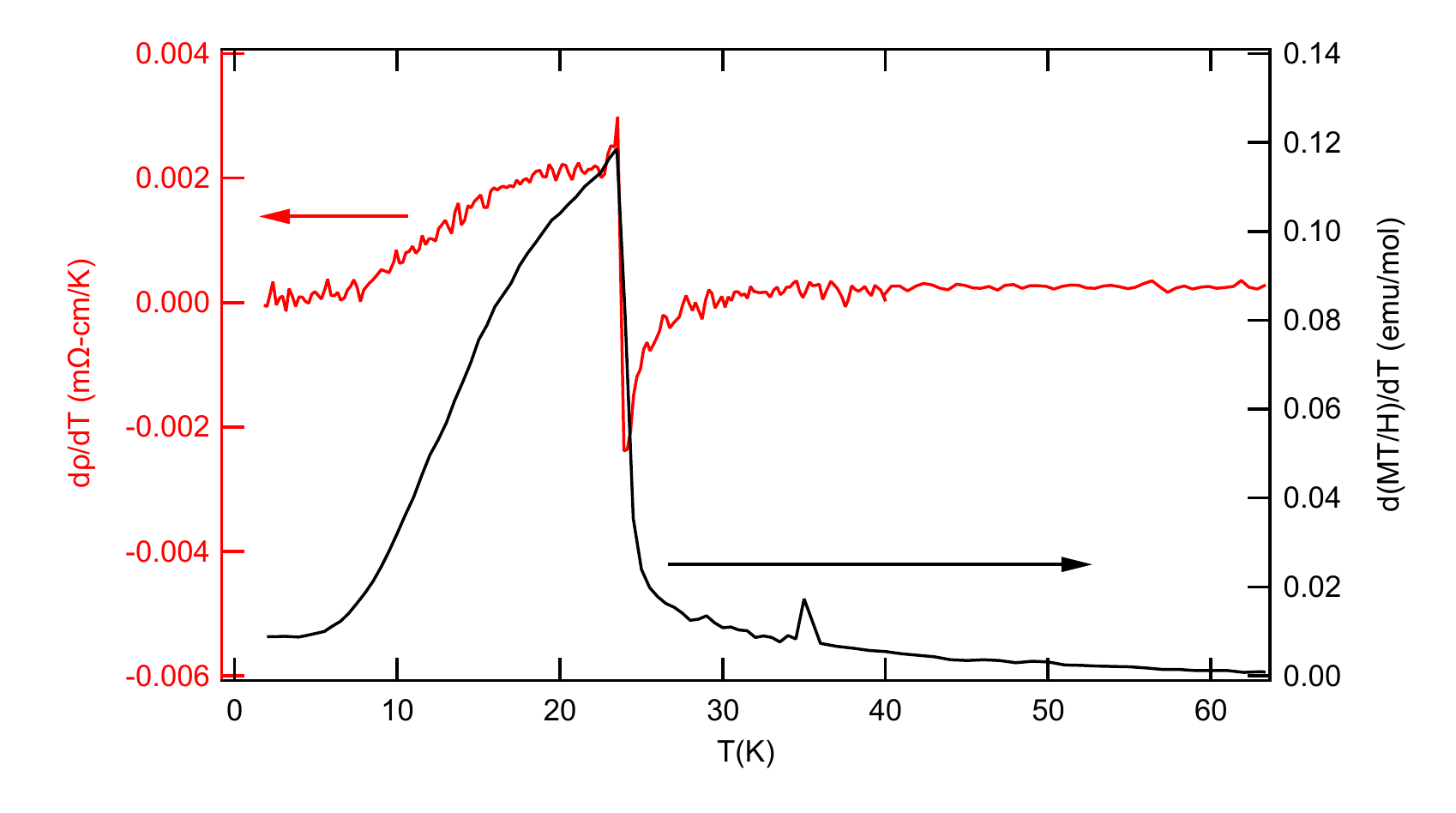}%
	\caption{
	Derivatives of resistivity and magnetization (d$\rho$/dT and d(M(T)/H)/dT) plotted as a function of temperature.
	\label{fig:derriv}}
\end{figure*}

\begin{figure*}[tb]
	\includegraphics[width=6 in]{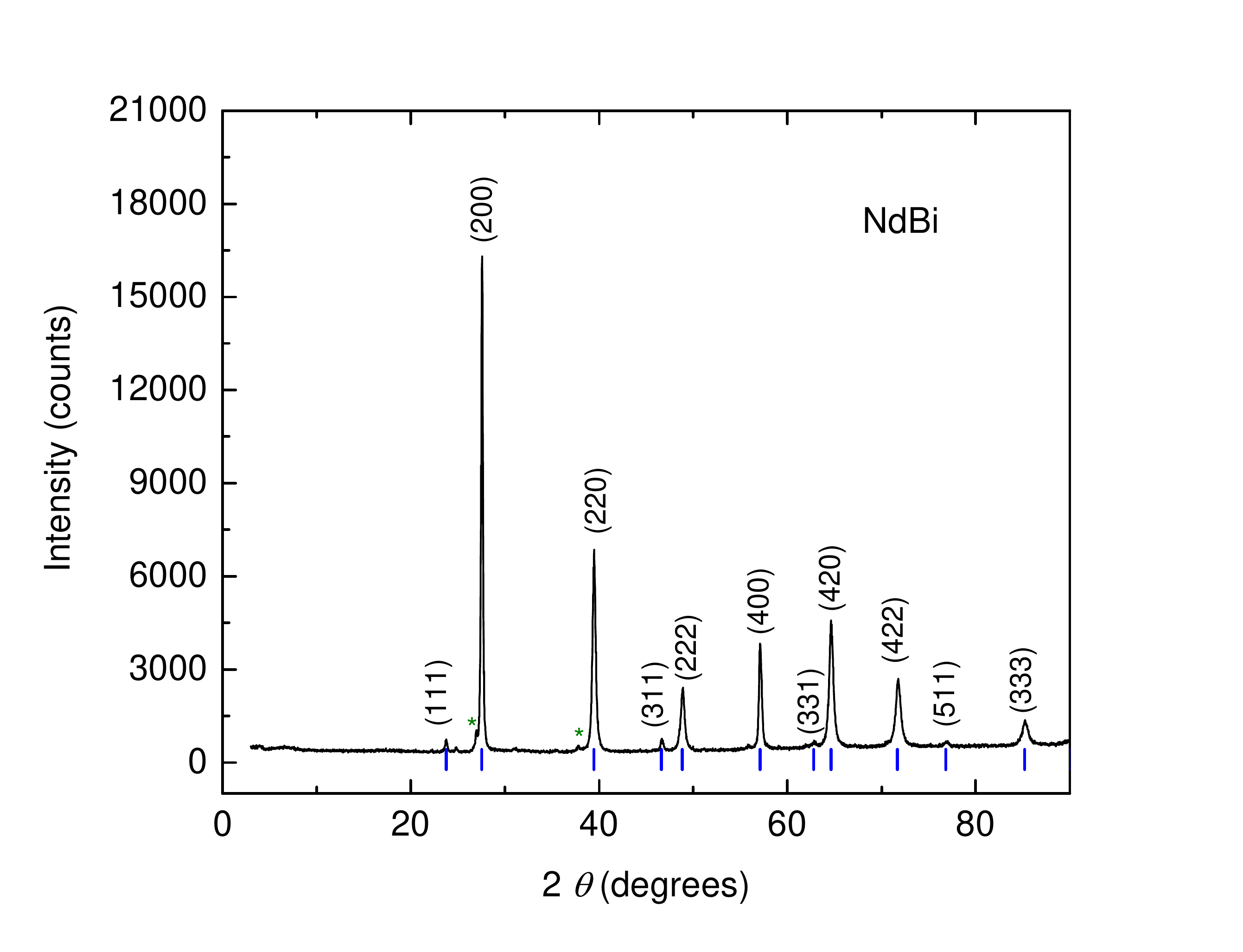}%
	\caption{
	Powder x-ray diffraction pattern for NdBi. Blue vertical lines show the expected Bragg peak positions, agreeing well with the observed pattern. Green asterisks show impurity peaks coming from Bi flux. hkl indices are assigned comparing with the reported structure. 
	\label{fig:xrd}}
\end{figure*}

\clearpage

\section{Photon energy dependent data}

To demonstrate that the states which appear below TN are indeed surface states, we present photon energy dependence of the band structure in this section.  All data was acquired in the AFM state at T = 4.5 K


In Supplementary Fig. S3, we present high statistic Fermi surface maps measured with different photon energy from 5.7 eV to 6.89 eV near the BZ part where new states appear in the AFM state. Near the Fermi surfaces plots, we show one cut along $\Gamma$-X direction and a cut in the direction orthogonal to it for each photon energy. No changes can be seen in the shape of the Fermi surface pockets nor in the corresponding dispersion plots. To show this precisely, we compared shapes of the disputations obtained by fitting energy distribution curves. The results are shown in Fig. S3. The curves match each other almost perfectly, well within experimental error bars. We note that for related compound YSb, the bulk band dispersion changes by more than 150 meV while tuning the photon energy from 5.5 eV to 6 eV \cite{Yun_PRB2017}. Therefore the data in Fig. S3 demonstrates the surface state character of the features that appear in AFM state of NdBi.

We demonstrate that these surface states exist only in the AFM state and disappear above T$_N$ also when measured using higher photon energies in Fig. S4 a, b. These features produce sharp sets of peaks in momentum distribution curves (MDCs) below T$_N$ that sit on top of broad tail of intensity from bulk bands (Fig. S4 c).  In the PM state, these peaks disappear and only broad tail of intensity due to bulk band projection. 

\begin{figure*}[tb]
	\includegraphics[width=3.5 in]{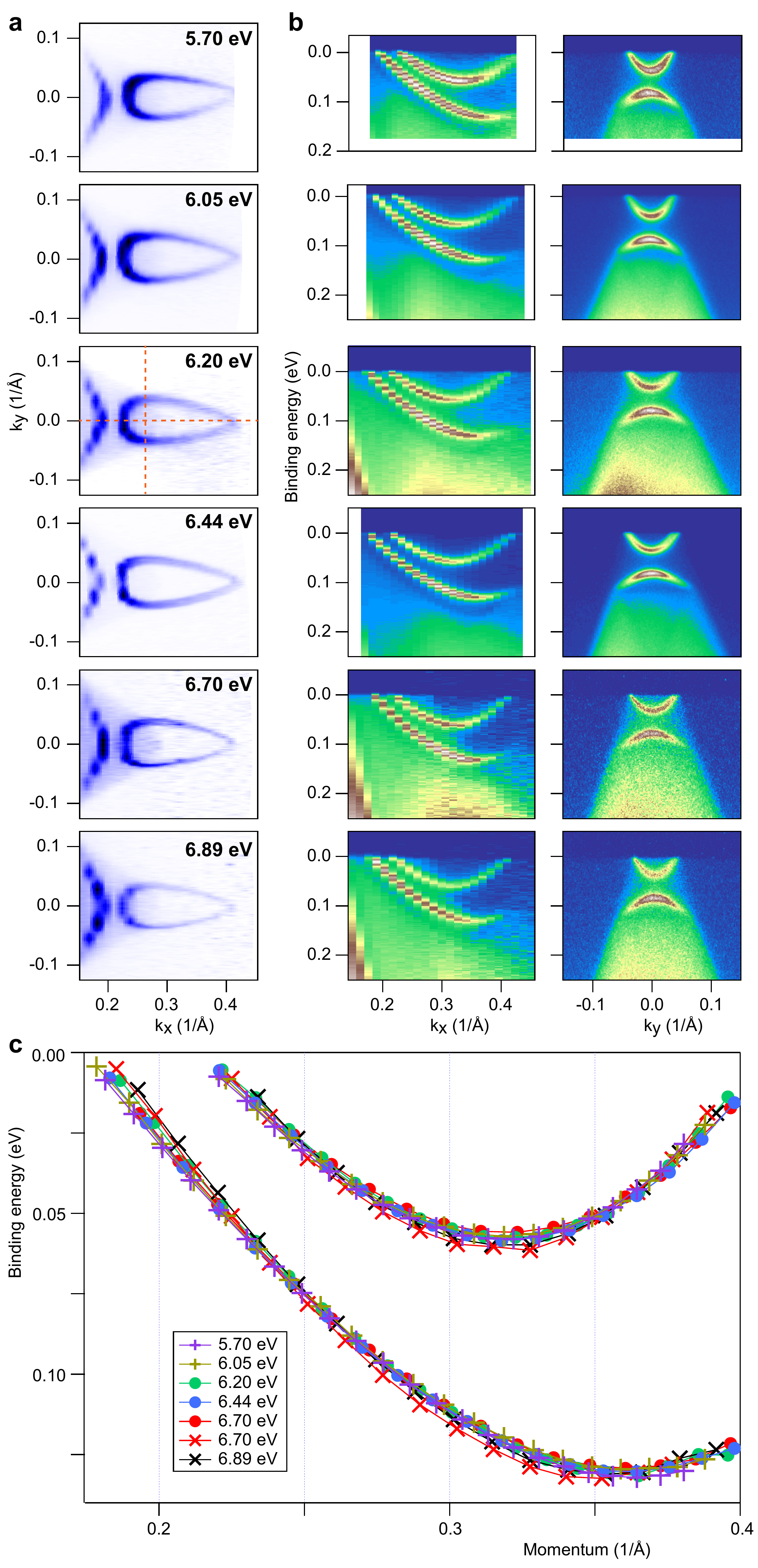}
	\caption{
	Photon energy dependent data. 
	{\bf a,} Fermi surface maps in the AFM state measured using light with different photon energy are shown in the left column. 
	{\bf b,} Corresponding dispersions along directions depicted by the horizontal and the vertical line on the map are shown in the central and the right columns, respectively. 
	{\bf c,} The shape of the dispersion in $\Gamma$-X direction obtained by EDC fitting. Bullets and crosses represent results obtained from two different cleaves.
	\label{fig:fig1}}
\end{figure*}

\begin{figure*}[tb]
	\includegraphics[width=6 in]{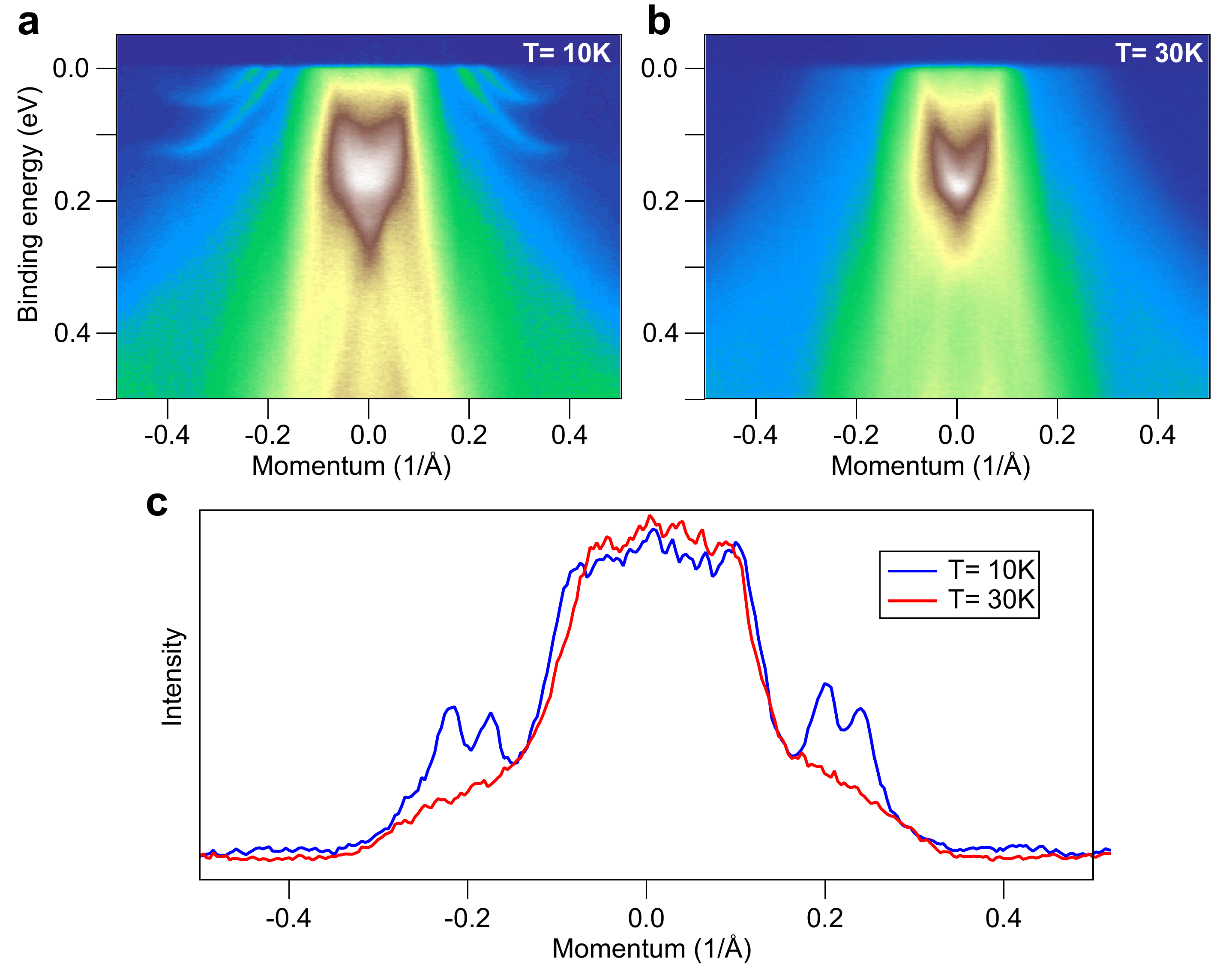}%
	\caption{
	ARPES data from NdBi in AFM and PM states measured using 21.2 eV photons. 
	{\bf a,} Data measured along $\Gamma$ - X direction in the AFM state at T=10~K.
	{\bf b,} Data measured along $\Gamma$ - X direction in the PM state at T=30~K.
	{\bf c,} Momentum Distribution Curves (MDC) plotted at E$_F$ from data in panels a, b.
	\label{fig:fig4}}
\end{figure*}
\clearpage

\section{Polarization dependent data}

Polarization of incident photons has significant effects on intensity in ARPES experiments. It provides a powerful tool to decipher the character and origin of the observed bands. Performing ARPES experiments using circular polarization are particularly useful as they can provide valuable information about the possible existence of spin textures \cite{GedikCD}. However this has to be carried out with care, as dichroic signal is expected to change sign across the experimental mirror plane (defined by sample normal and direction of incoming photon beam) even for trivial and core level states due to "handedness" of photoemission process defined by vectors of incoming beam, sample normal and directions of outgoing photoelectrons \cite{Venus}. Spin textures, such as present in topological insulators, manifest themselves by having opposite sign change across the mirror plane for the upper and lower parts of the Dirac cone \cite{GedikCD, ScholzCD} as well as change of sign within the same band at momenta away from the mirror plane \cite{GedikCD}. 

We can confirm that the hole and electron surface states reported in this work have opposite dichroic response, where the sign of the CD signal for the two bands changes across the experimental mirror plane in an opposing way. This is illustrated in Supplementary Fig. S5 d, e. In addition, upon sample rotation by 45 deg, the sign of the dichroic signal changes within the same band at momenta away from the mirror plane, as evident in Supplementary Fig. S5 f. This is consistent with presence of opposite spin textures in hole and electron surface states as demonstrated previously \cite{GedikCD}.
We contrast this behavior with dichroism signal measured above T$_N$ in the PM state and shown in Fig. S5 i, j, where such effects are not observed.

The beam with circularly left  (LCP), circularly right (RCP), linear horizontal (LH) or linear vertical (LH) polarization was produced using zero-order tunable phase retardation plate from Alphalas GmbH. The tilt of the plate was adjusted using a calibration curve and full circular polarization was verified using external polarimeter. The location of the beam at sample position was recorded for different polarization using phosphorescent plate, and movement of the beam during switching of polarization was corrected using beam deflector within half of the beam size. 

\begin{figure*}[tb]
	\includegraphics[width=7 in]{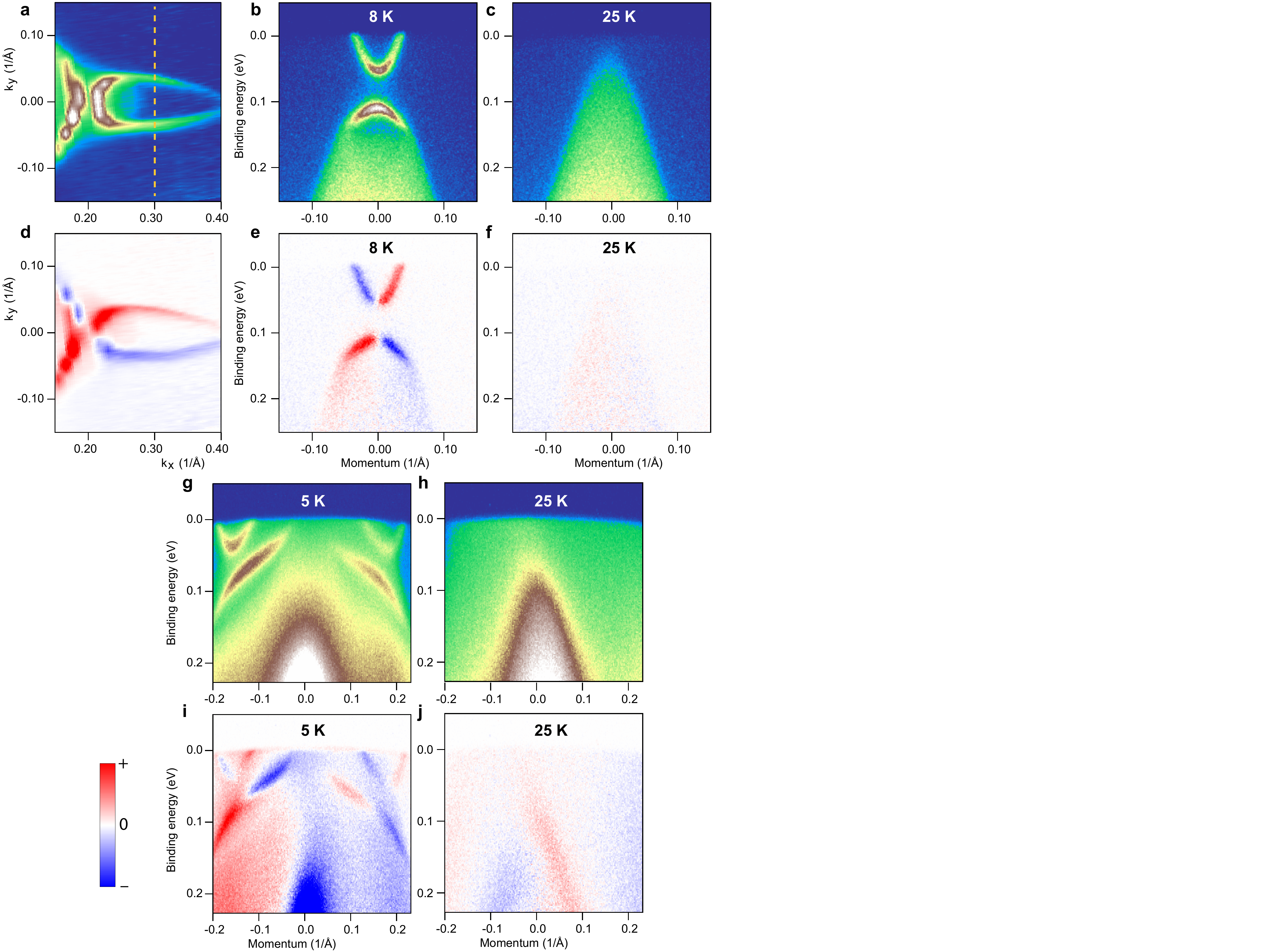}%
	\caption{
	Circular dichroism in ARPES data.
	{\bf a,} Plot of portion of Fermi surface in the area where surface states are present measured using linearly polarized light
	{\bf b,} and {\bf c,} ARPES data along cut marked in panel (a) measured in the AFM and PM state respectively (sum of spectra measured using left and right circular polarisation).
	{\bf d,} dichroic signal from FS as in panel (a)
	{\bf e,} and {\bf f,} dichroic signal in AFM and PM state along same cuts as in (b) and (c) (difference of spectra measured using left and right circular polarisation).
	{\bf g,} and {\bf h,}  ARPES data along cut \#5 marked Fig. 3c in 45 deg. sample orientation in the AFM and PM state respectively.
	{\bf i,} and {\bf j,} dichroic signal in AFM and PM state along same cuts as in (g) and (h)
	\label{fig:fig2}}
\end{figure*}

\begin{figure*}[tb]
	\includegraphics[width=5 in]{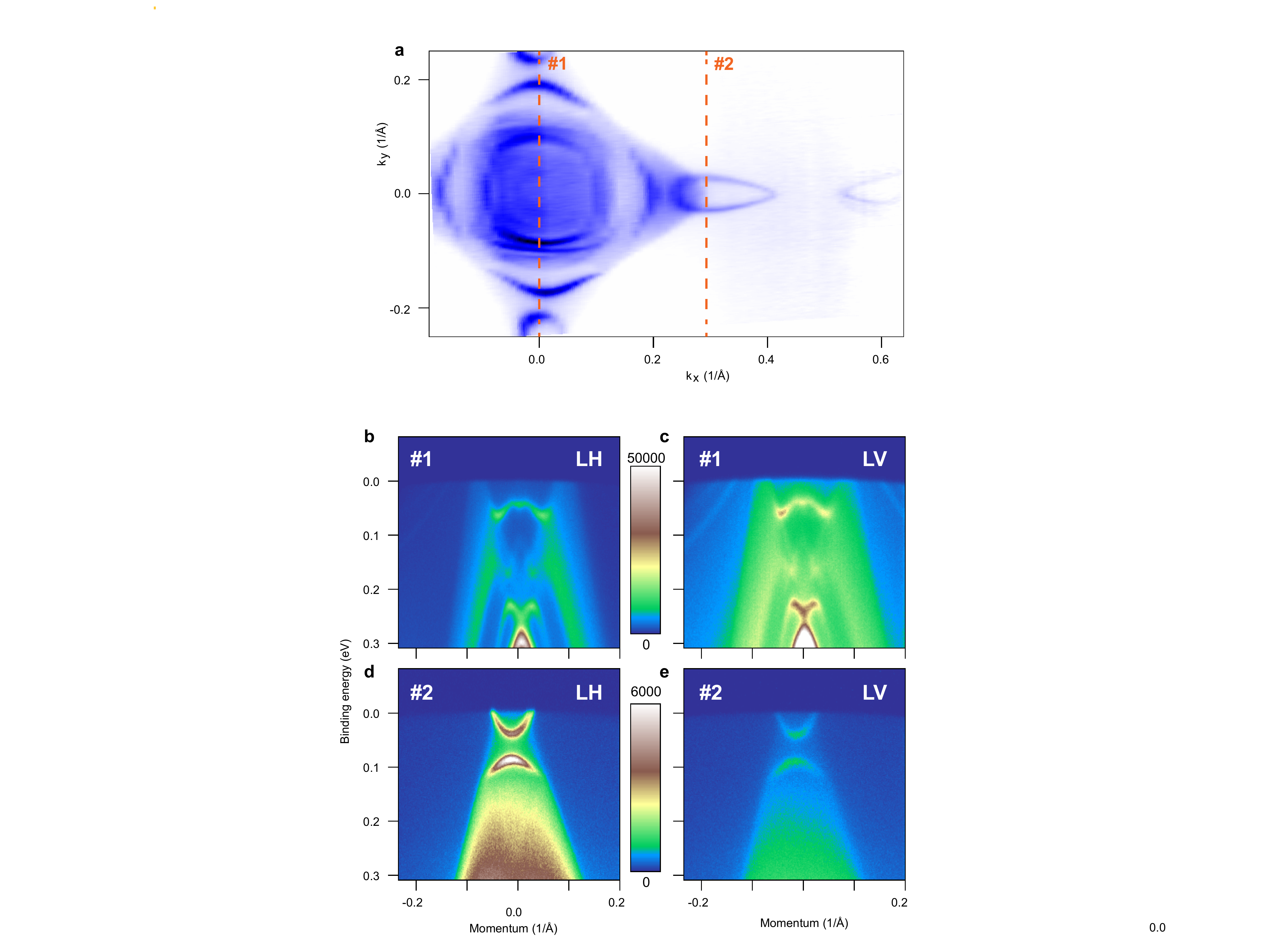}%
	\caption{
	ARPES data acquired with horizontal and vertical polarizatrions. 
	{\bf a} Fermi surface map.
	{\bf b} and {\bf c,} ARPES spectra along cut marked with line \#1 in (a) measured using linear horizontal and linear vertical polarized light, respectively.
	{\bf d} and {\bf e,} ARPES spectra along cut marked with line \#2 in (a) measured using linear horizontal and linear vertical polarized light, respectively.
	\label{fig:fig2}}
\end{figure*}

\clearpage

\section{Reproducibility study}

To show the reproducibility of surface states, we show Fermi surface maps obtained from eight different samples in Supplementary Fig. S7a. Maps \#1-3 were measured in the experimental geometry in which the analyzer silt is parallel $\Gamma-M$ direction, and maps \#4-8 were measured in the experimental geometry in which the analyzer silt is parallel $\Gamma-X$ direction. All maps demonstrate the presence of the surface state electron-like pocket and the arc.

In Supplementary Fig. S7b, We demonstrate how the surface state recovers after heating the sample up to $T > T_N$ and cooling it back down. The started measurements at 6K; at this temperature, surface states are clearly seen. We gradually increased the temperature to 25K; at this temperature, the sample becomes paramagnetic, and the surface states are absent. After cooling down, the surface states recovered. The spectra measured at T=15K and T=20K are similar to spectra measured at these temperatures before.

Also, these data show that aging effects are weak, and the surface states can survive for a long time during an experiment. At least, it is true for the experimental conditions of our laser ARPES setup. Another example of a long experiment is preset in Fig. S7c. Both datasets in Supplementary Fig. S7b, c show that aging effects (primarily band broadening) can become noticeable during a long experience, but even after 24 hours, all bands are clearly visible.

\begin{figure*}[tb]
	\includegraphics[width=5 in]{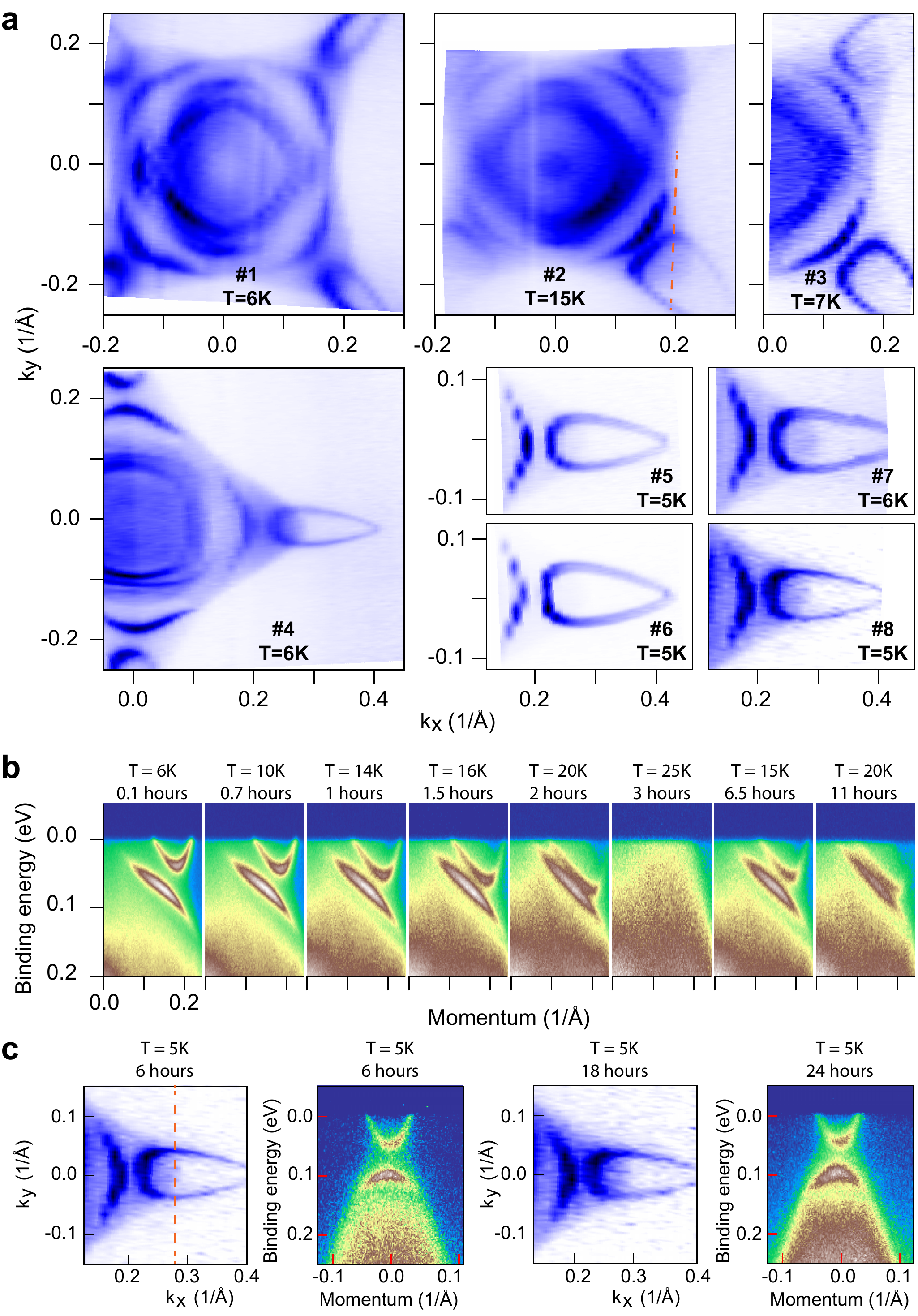}%
	\caption{
	Reproducibility and aging data. 
	{\bf a,} Fermi surface maps measured from eight different samples in the AFM state.
	{\bf b,} Spectra measured along the direction shown with a dashed line in the plot {\bf a}\#2 at different temperatures. A number of hours represents the time after cleave when a spectrum was measured.
	{\bf c,} Two Fermi surface maps and two spectra measured from another sample at different times after cleave: from 6 to 24 hours. The spectra were measured in along direction shown with a dashed line in the map.
	\label{fig:fig3}}
\end{figure*}
\clearpage

\section{Synchrotron data}

Data measured using higher photon energy at synchrotron is shown in Supplementary Fig. 8. The Fermi surface map reveals the same features as measured with VUV laser - surface state Fermi arcs and electron pockets that are present only below T$_N$. The sample aging was more severe at the synchrotron due to worse UHV condition. This causes downward shift of the E$_F$ indicating removal of electron from the sample likely due to condensation of CO. These effects were mostly absent when performing measurements in the lab using laser source as demonstrated in previous section.

\begin{figure*}[tb]
	\includegraphics[width=5 in]{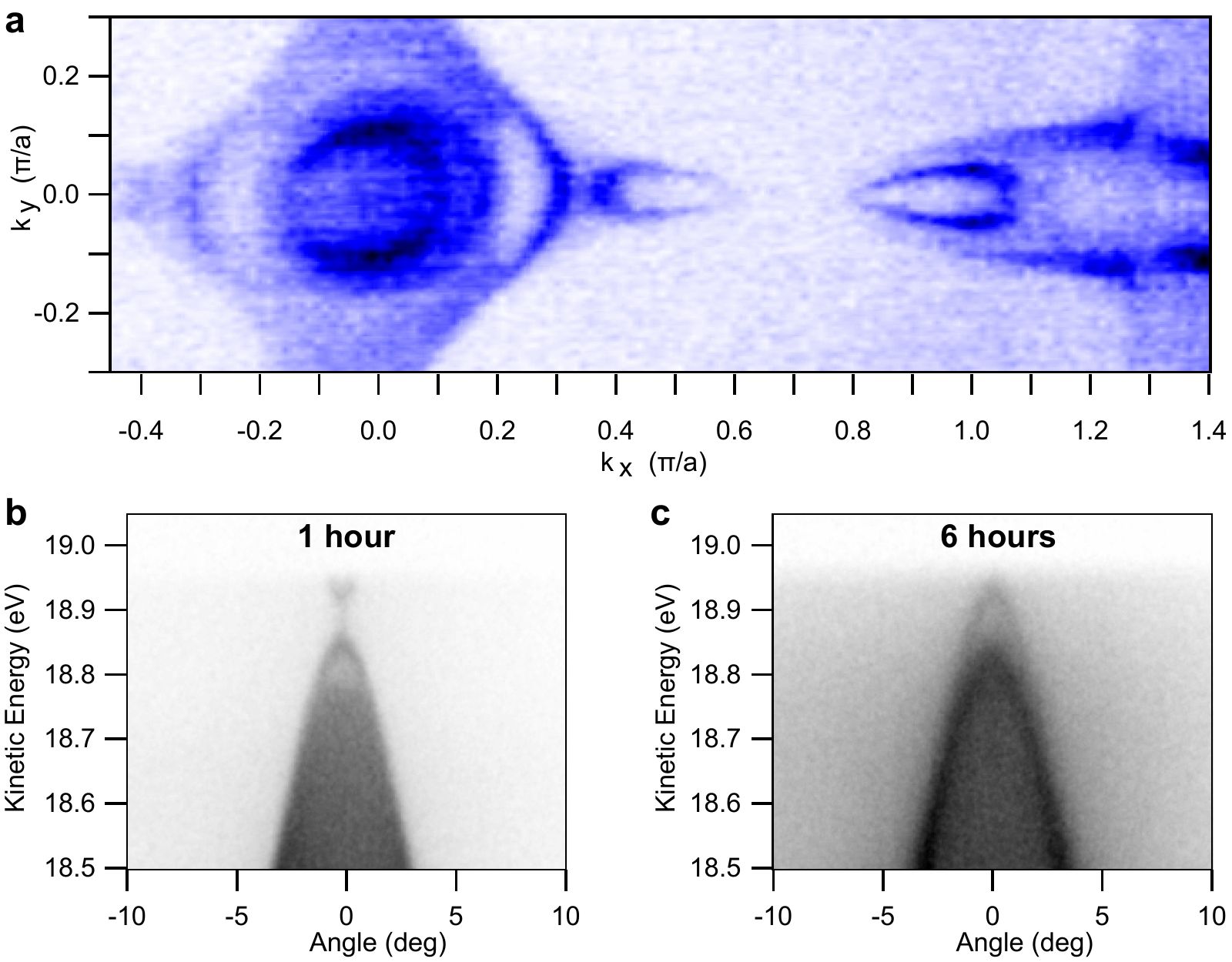}%
	\caption{
	Synchrotron data.
	{\bf a,} Fermi surface map measured using 23eV light at T=8K.
	{\bf b,} Spectrum measured throw the surface electron-like pocket on another sample one hour after its cleaving.
	{\bf c,} The same spectrum measured 5 hours later.
	\label{fig:fig4}}
\end{figure*}
\clearpage

\section{Emergence of  Fermi arcs and novel magnetic splitting in an antiferromagnet}

The long range AFM order doubles the size of real space unit cell, and therefore introduces magnetic zone boundary within the Brillouin zone. The small periodic potential creates shadow bands, which are weak mirror copies of the bands about the magnetic zone boundaries (MZB). The magnetic ordering in NdBi is of AFM A type, which consists of planes that have the same orientation of magnetic moments that are stacked in AFM fashion along one axis. 

\begin{figure*}[tb]
	\includegraphics[width=6 in]{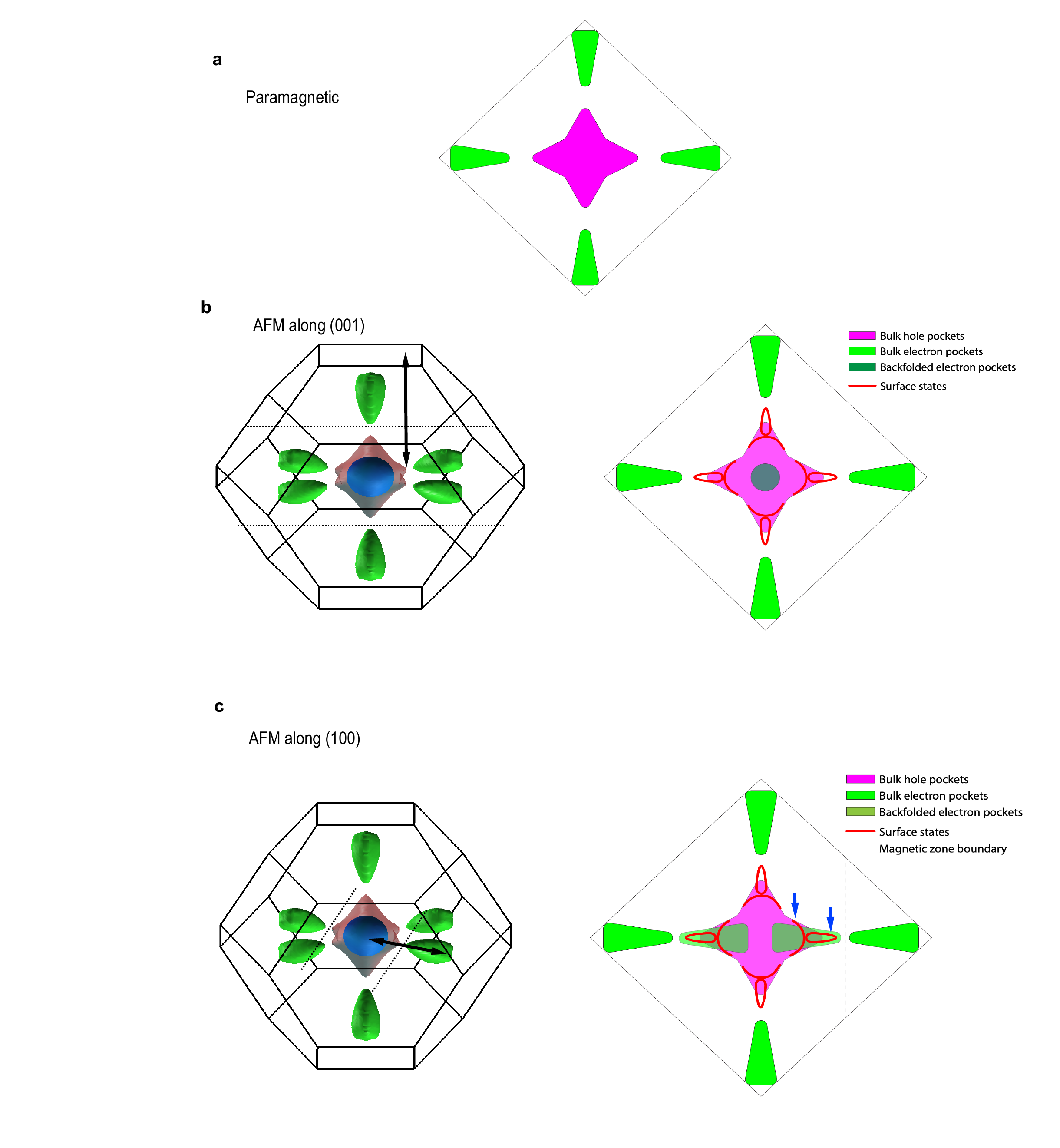}%
	\caption{
	Schematic illustration of various band backfolding scenarios.
	{\bf a,} schematic plot of 2D projected bulk FS based on calculations from Fig. 1c.
	{\bf b,} 3D bulk FS and 2D projection when magnetic ordering vector is perpendicular to sample surface. 
	{\bf c,} same as b, but for magnetic order parallel to the sample surface. Blue arrows point to parts of the surface states that exist outside the areas where bulk electron shadow band overlaps with bulk hole pocket.
	\label{fig:foldingschem}}
\end{figure*}

If the AFM ordering vector is along the direction perpendicular to the sample surface i. e. along (001) direction, then MZB are horizontal planes (as illustrated in Supplementary Fig. S9b. In such case, the top and bottom electron pocket are back folded onto hole pockets at Gamma and vice versa. Since there are no states near E$_F$ that would be backfolded onto areas, where we observe the surface states, such scenario cannot be responsible for their emergence.

\begin{figure*}[tb]
	\includegraphics[width=4 in]{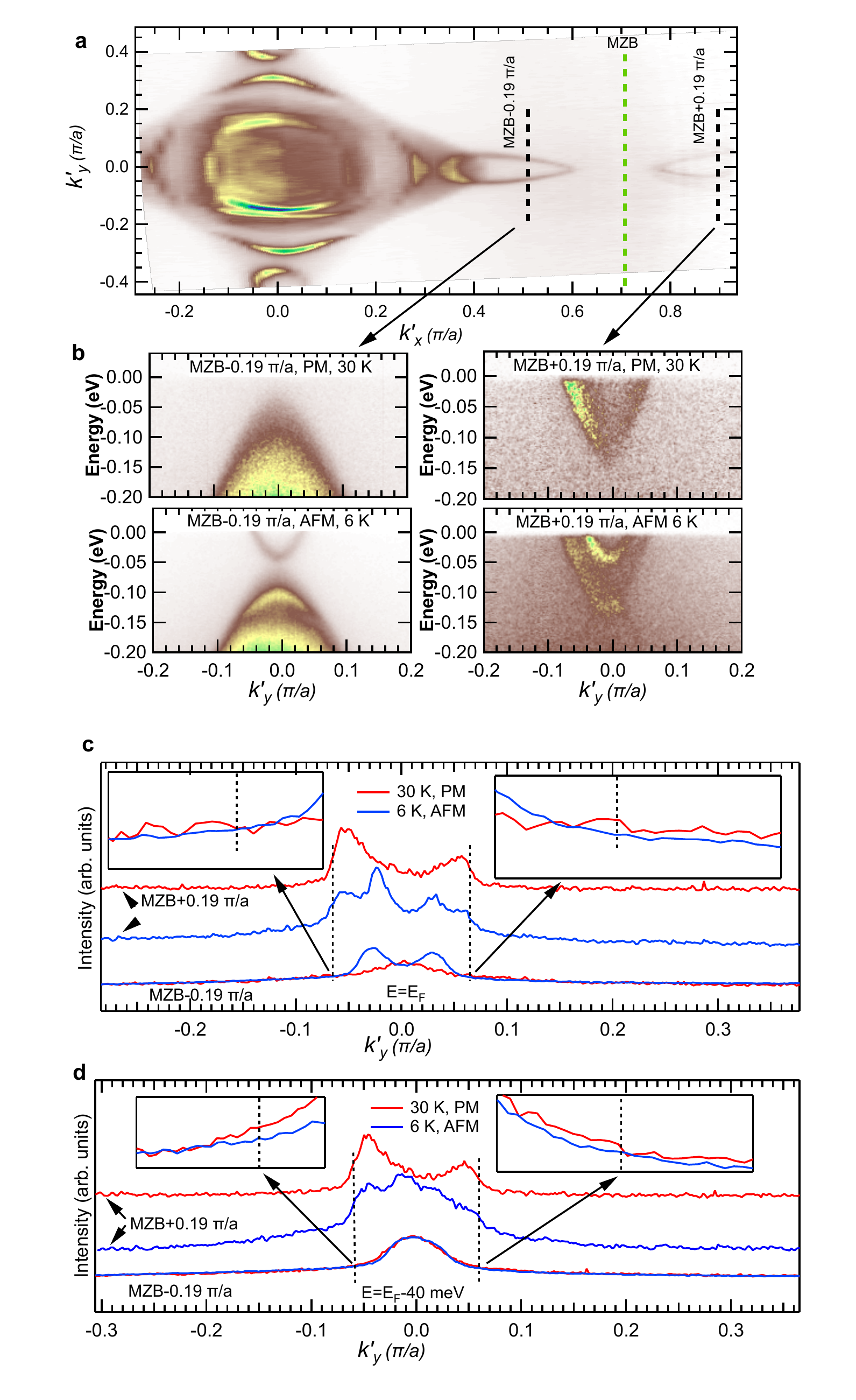}%
	\caption{
	Absence of observable signal from bulk shadow bands.
	{\bf a,} FS plot in AFM state. Green dashed line marks MZB for in plane magnetic ordering vector.
	{\bf b,} data along cuts on left and right side, equally distanced from MZB in PM state (top row) and AFM state (bottom row).
	{\bf c,} MDC from cuts in panel b at E$_F$. Dashed lines mark strong shoulders from bulk electron pocket on right side of MZB. Insets show magnified data at momenta where bulk electron shoulders due to shadow bands are expected.
	{\bf d,} MDC from cuts in panel b at E=-40 meV. Dashed lines mark strong shoulders from bulk electron pocket on right side of MZB. Insets show magnified data at momenta where bulk electron shoulders due to shadow bands are expected.
	\label{fig:foldingdata}}
\end{figure*}

More interesting case happens, when the AFM ordering vector is parallel to the sample surface. Then there could be a potential overlap between central hole  pocket and shadow bands of the electron pockets in the general area where surface states appear. This is schematically illustrated in Supplementary Figure S9c. Even then, the surface states we observe exist outside of the areas of the overlap between hole and electron shadow bands, as indicated by blue arrows, therefore they cannot be a result of hybridization due to band back folding. 

We can also demonstrate that the back folding effects of the bulk bands are very weak in this material and below our detection limit. In Supplementary Fig. S10 we plot the data along cuts that are equally spaced on left and right side of the MZB. The back folding in AFM state should produce shadow bands and features from right side should appear in data on the left side and vice versa. We note that the bulk electron band from right PM cut (right top of panel b) , should appear with diminished intensity as shadow band overlap the PM data on the left cut (top left of panel b). The actual data in AFM state on the left side, does not show even a hint of such intensity that should exist below surface electron band. We illustrate this in more detail by plotting MDC's at E$_F$ and -40 meV in panels c and d. The bulk electron pocket produces strong shoulders marked by dashed lines. In presence of strong band folding those shoulders should appear in the AFM MDC's from the left side of the MZB. Magnified data shows absence of such shoulders down to noise level, which puts upper limit on intensity of shadow bands of $\le$ 3 \%. This makes the band backfolding very unlikely to be responsible for emergence of the surface state Fermi arcs that have intensity that is comparable to intensity of PM bulk bands.

We illustrate difficulties with explaining the emergence of surface states in AFM state using the back folding scenario by simulating its effects and comparing to actual data in Supplementary Fig. S11. In top two panels we plot the data in PM state along the same  cuts on left and right side of MZB as in Supplementary Fig. S10b. In panel c we plot data from panel a, after adding 10 \% of intensity from panel b to simulate 10 \% shadow band intensity. The result has clearly visible features indicated by an arrow that are missing from actual data in AFM state shown in panel d. The surface state electron pocket present in AFM state is not only much more intense, but also is located at different binding energy than the potential shadow band.

\begin{figure*}[tb]
	\includegraphics[width=5 in]{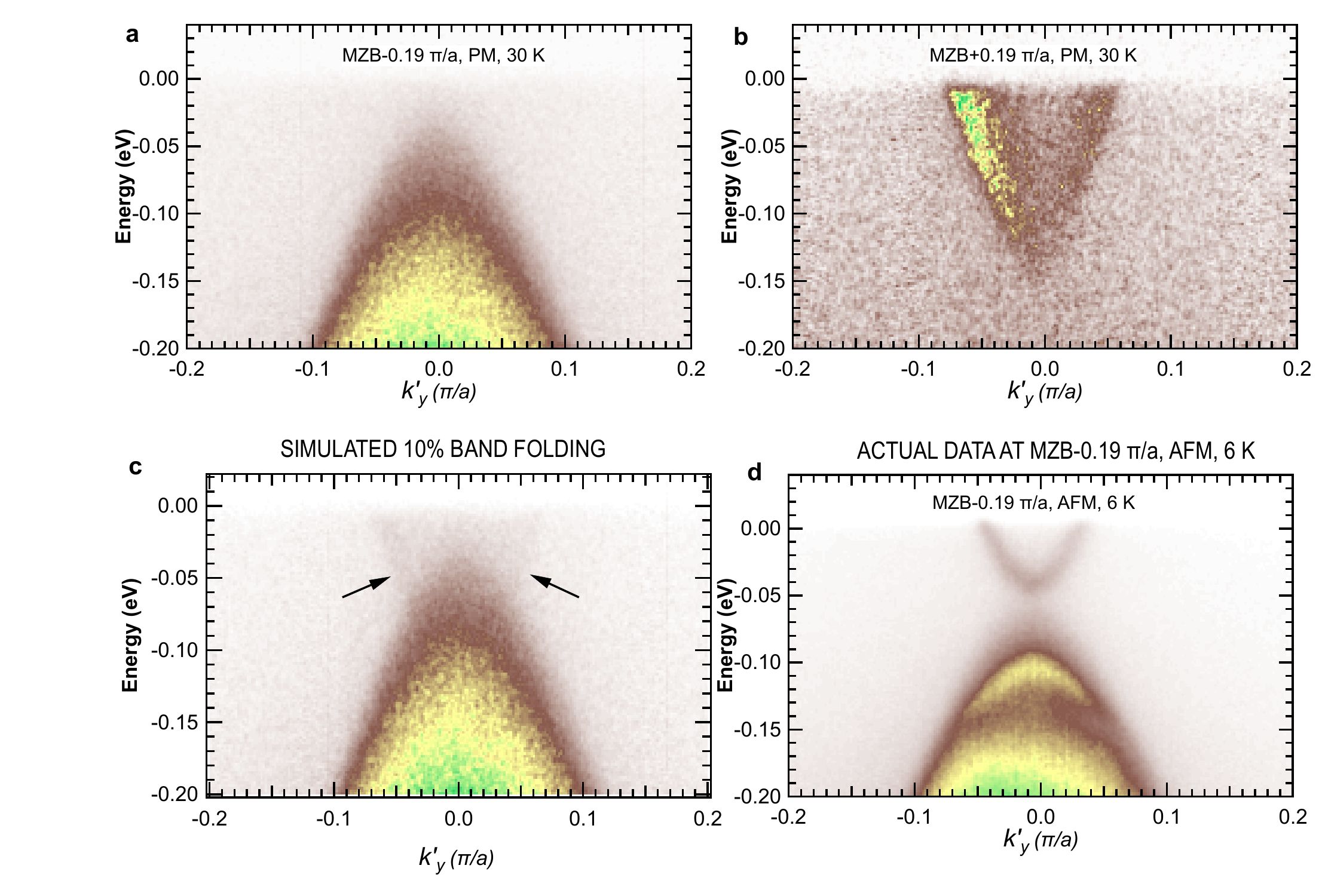}%
	\caption{
	Simulation of shadow intensity for in-plane backfolding. 
	{\bf a,} Data along the cut on the left side of MZB (same as Fig. S9b). 
	{\bf b,} Data along the cut on the right side of MZB (same as Fig. S9b). 
	{\bf c,} Sum of data in panel a and 10 \% of intensity from panel b simulating 10 \% backfolding in AFM state. 
	{\bf d,} Actual data along the same cut as in panel a in AFM state. 
	\label{fig:foldingsim}}
\end{figure*}

\clearpage

\section{DFT band structure calculations}

For trying to understand the exotic magnetic Fermi arc and surface states observed by ARPES in NdBi below $T_N$, we have carried out surface band structure calculations of different magnetic configurations based on DFT as shown in Supplementary Fig.S12, but so far none of them can explain the observed magnetic Fermi arc. First for the paramagnetic or non-magnetic (NM) state above $T_N$, the NdBi (001) surface spectral function in  Fig.S12(a) shows the band inversion along $\overline{\Gamma}$-$\overline{M}$ as projected from the avoided crossing with $p-d$ band inversion in the $\Gamma$-X direction of the bulk Brillouin zone (BZ). Such band inversion makes the NM NdBi a strong topological insulator with the topological index of $Z_2$=(1;000), which is typical for NM early rare-earth (R) bismuth 1:1 compounds in the same rock salt structure \cite{zeng2015topological, wu2016asymmetric, DuanCommPhys2018}. Because of two bulk X points with the band inversion being projected on the same $\overline{M}$ point, two surface Dirac cones overlap and are gaped to form an upper and lower surface band crossing at the $\overline{M}$ point, which also agrees with earlier studies on paramagnetic RBi systems \cite{zeng2015topological, wu2016asymmetric, Kuroda_2018, li2018tunable}. For the 2D Fermi surface (FS) at E$_F$ in Fig.S12(d), in comparison to the bulk-only FS in Fig.1C, the surface states around the $\overline{M}$ point connect and merge with the bulk electron pocket along the $\overline{\Gamma}$-$\overline{M}$ direction. At lower energy of E$_F$ -- 0.15 eV (Fig.S12(g)), these surface states form an envelope around the shrinking bulk electron pocket.

For the anti-ferromagnetic (AFM) configures below $T_N$, we used the A-type AFM (AFMA) stacked along (001) with moment also along (001) as reported in the early neutron study by Nereson and Arnold \cite{nereson_1971}. Our DFT+U+SOC calculation with U=6.3 eV and J=0.7 eV gives a spin moment of 2.7 $\mu_B$ and an orbital moment of 5.8 $\mu_B$ in the opposite direction, resulting in a total magnetic moment of 3.1 $\mu_B$ on Nd, agreeing with the experimental data of 3.1$\pm$0.2 $\mu_B$ \cite{nereson_1971}. Although time-reversal symmetry (TRS) is broken, the non-symmorphic TRS exists and when combined with inversion symmetry, the bulk bands still have double degeneracy. The calculated topological index $Z_4$=2 and there is also a bulk Dirac point at E$_F$ + 0.4 eV being projected at the $\overline{\Gamma}$ point. For surface states, as shown in Fig.S12(b), the empty Nd 4$f$ bands are 0.5 eV above the E$_F$. The band inversion along $\overline{\Gamma}$-$\overline{M}$ still holds, but the two surface band crossings at the $\overline{M}$ point are both gaped out. In contrast to the hole Fermi arc and two surface electron pockets in the ARPES data in Fig.2E, Fig.S12(b) from the calculated AFMA shows no surface states extending beyond the band inversion region along the $\overline{\Gamma}$-$\overline{M}$ direction toward near the $\overline{\Gamma}$ point to merge into the bulk band projections. The 2D FS of AFMA at different energies in Fig.S12(e) and (h) are similar to those in the NM case and do not match that of ARPES data in Fig.2D because of clearly missing the hole Fermi arc and the surface electron pocket on the $\overline{\Gamma}$ point side.

To evaluate the band-folding effect, we also considered the intralayer checkerboard AFM (AFMC) configuration with moment along (001) as shown in Fig.S12 (c), (f) and (i). For AFMC, the (001) surface BZ (SBZ) as drawn in Fig.S12(e) is rotated by 45 degree and folded in reference to that of NM and AFMA. As shown in Fig.S12(c), the band inversion along the original $\overline{\Gamma}$-$\overline{M}$ is folded at the half-point and on top of new $\overline{M'}$-point. The surface states of the overlapped Dirac cones at the original $\overline{M}$-point are folded on top and centered at the $\overline{\Gamma}$ point. However, the surface states and the 2D FS at different energies in Fig.S12 (f) and (i) do not match the surface hole Fermi arc and two surface electron pockets in ARPES that appear along the $\overline{\Gamma}$-$\overline{M}$ on top of the well separated bulk hole and electron pockets in Fig.2E. For AFMC, the bulk electron pockets with the associated surface states are folded directly on top of the bulk hole pocket centered at the $\overline{\Gamma}$ point. Also notably the $C_4$ rotational symmetry is broken because the stacking of the checkerboard AFM needs an in-plane non-symmorphic translation.

\clearpage
\newpage
\clearpage

\begin{figure*}[tb]
	\includegraphics[width=5 in]{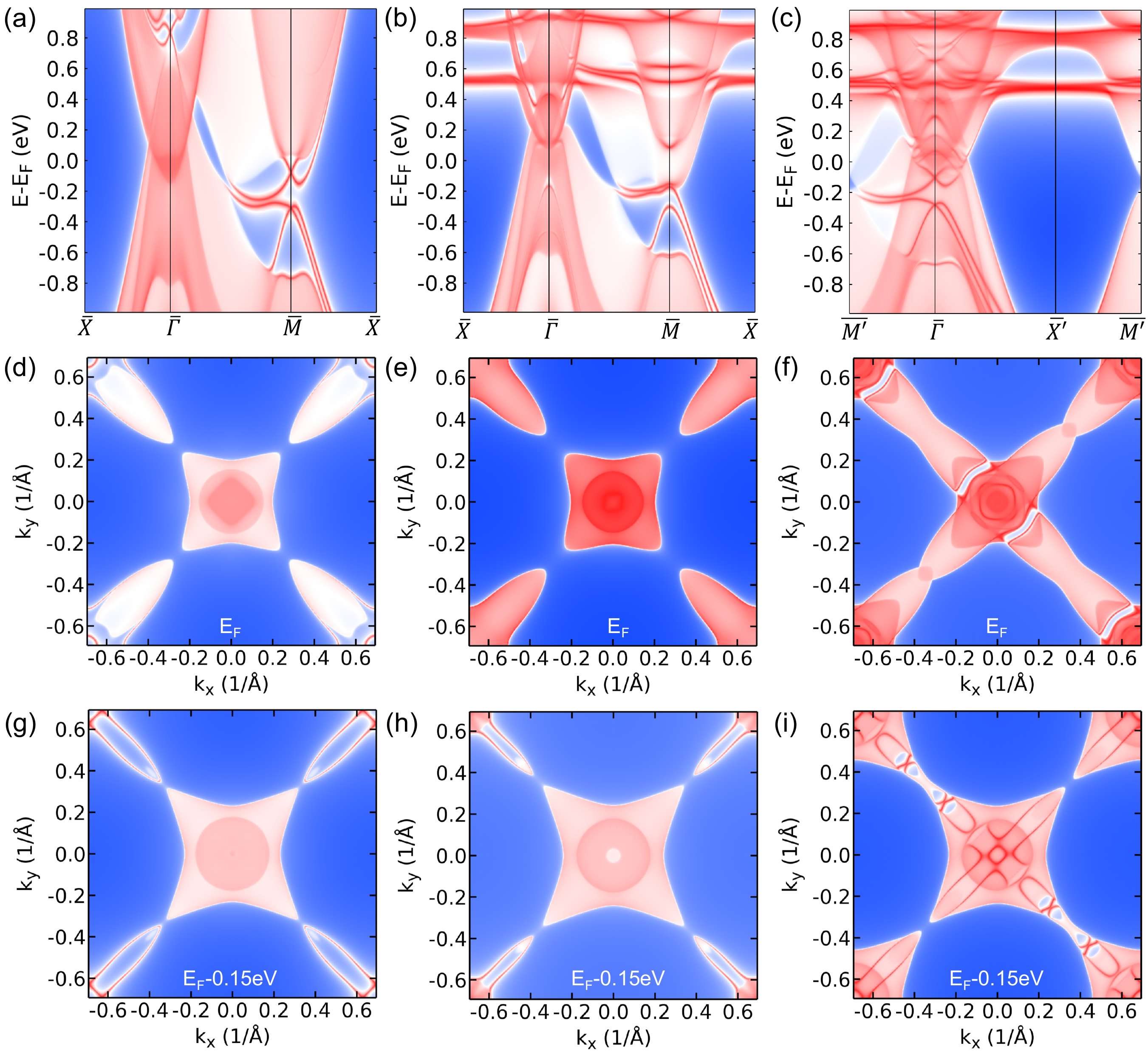}%
	\caption{
	Spectral functions of NdBi (001) surface in (a) paramagnetic or non-magnetic (NM) configuration, (b) A-type anti-ferromagnetic (AFMA) configuration stacked along (001) with moment also along (001) and (c) intralayer checkerboard AFM (AFMC) with moment along (001). The 2D Fermi surfaces at the constant energy of E$_F$ ((d), (e) and (f)) and E$_F$–0.15 eV ((g), (h) and (i)) for the three magnetic configurations (NM, AFMA and AFMC), respectively. The orange square in (e) shows the surface Brillouin zone of AFMC as rotated by 45 degree and folded from that of NM and AFMA.
	\label{fig:fig5}}
\end{figure*}

\end{document}